\documentclass[journal]{IEEEtran}
\usepackage{times}
\usepackage{epsfig}
\usepackage{graphicx}
\usepackage{amsmath}
\usepackage{amssymb}
\usepackage[ruled]{algorithm2e}
\usepackage{algpseudocode}
\usepackage{caption}
\usepackage{subfigure}
\usepackage{booktabs}
\usepackage{tabularx}
\usepackage{xcolor}
\usepackage{multirow}
\usepackage{hyperref}

\newcolumntype{Y}{>{\centering\arraybackslash}X}

\begin{document}

\title{AutoFi: Towards Automatic WiFi Human Sensing via Geometric Self-Supervised Learning}

\author{Jianfei~Yang,
	Xinyan~Chen,
	Han~Zou,
	Dazhuo~Wang,
 	and~Lihua~Xie,~\IEEEmembership{Fellow,~IEEE}
 
 \thanks{
	J. Yang, X. Chen, D. Wang and L. Xie are with the School of Electrical and Electronics Engineering, Nanyang Technological University, Singapore (e-mail: yang0478@ntu.edu.sg; elhxie@ntu.edu.sg).
 
 	H. Zou is with the Department of Electrical Engineering and Computer Sciences, University of California, Berkeley, USA (e-mail: enthalpyzou@gmail.com).
	
	This work is supported by NTU Presidential Postdoctoral Fellowship, ``Adaptive Multimodal Learning for Robust Sensing and Recognition in Smart Cities'' project fund, in Nanyang Technological University, Singapore.
	}
}

\markboth{IEEE Internet of Things Journal}%
{Shell \MakeLowercase{\textit{et al.}}: Bare Demo of IEEEtran.cls for IEEE Journals}

\maketitle

\begin{abstract}
   WiFi sensing technology has shown superiority in smart homes among various sensors for its cost-effective and privacy-preserving merits. It is empowered by Channel State Information (CSI) extracted from WiFi signals and advanced machine learning models to analyze motion patterns in CSI. Many learning-based models have been proposed for kinds of applications, but they severely suffer from environmental dependency. Though domain adaptation methods have been proposed to tackle this issue, it is not practical to collect high-quality, well-segmented and balanced CSI samples in a new environment for adaptation algorithms, but randomly-captured CSI samples can be easily collected. {\color{black}In this paper, we firstly explore how to learn a robust model from these low-quality CSI samples, and propose AutoFi, an annotation-efficient WiFi sensing model based on a novel geometric self-supervised learning algorithm.} The AutoFi fully utilizes unlabeled low-quality CSI samples that are captured randomly, and then transfers the knowledge to specific tasks defined by users, which is the first work to achieve cross-task transfer in WiFi sensing. The AutoFi is implemented on a pair of Atheros WiFi APs for evaluation. The AutoFi transfers knowledge from randomly collected CSI samples into human gait recognition and achieves state-of-the-art performance. Furthermore, we simulate cross-task transfer using public datasets to further demonstrate its capacity for cross-task learning. For the UT-HAR and Widar datasets, the AutoFi achieves satisfactory results on activity recognition and gesture recognition without any prior training. We believe that the AutoFi takes a huge step toward automatic WiFi sensing without any developer engagement. Our codes have been included in \href{https://github.com/xyanchen/WiFi-CSI-Sensing-Benchmark}{https://github.com/xyanchen/WiFi-CSI-Sensing-Benchmark}.
\end{abstract}

% Note that keywords are not normally used for peer-review papers.

%%%%%%%%% Structure
% 目前的wifi感知的机器学习方法都需要训练模型。训练模型后，对于不同环境，模型性能会有下降，很多工作提出可以用DA来解决，但是DA依然需要无监督地收集大量相对均衡的高质量样本，这即使是在客户端也是不可行的。

%%%%%%%%% BODY TEXT
\section{Introduction}
% WiFi sensing
\IEEEPARstart{W}{ith} the increasing demands of internet access, WiFi infrastructures have been ubiquitous and many mobile devices are equipped with WiFi modules. Multiple-Input Multiple-Output (MIMO) with Orthogonal Frequency-Division Multiplexing (OFDM) was innovated for higher requirement of data traffic in wireless communications~\cite{love2003grassmannian}. Along with very high spectral efficiency, MIMO provides the Channel State Information (CSI) for antenna pairs between receiver and transmitter devices. The CSI data records the propagation quality of multi-path wireless signals in specific environments, and therefore it enables WiFi-based radar technology~\cite{yang2013rssi,yang2018device}. WiFi-based radar can sense human motions by extracting CSI patterns by signal processing~\cite{wang2016human} or data-driven models~\cite{zou2018deepsense}, which has empowered many applications at smart homes including occupancy estimation~\cite{zou2018device}, activity recognition~\cite{yang2021deep}, gesture recognition~\cite{yang2019learning,zou2018robust}, human identification~\cite{zou2018identification}, human pose estimation~\cite{yang2022metafi} and vital sign detection~\cite{wang2016human}.

\begin{figure}[t]
	\centering
	\includegraphics[width=1.0\linewidth]{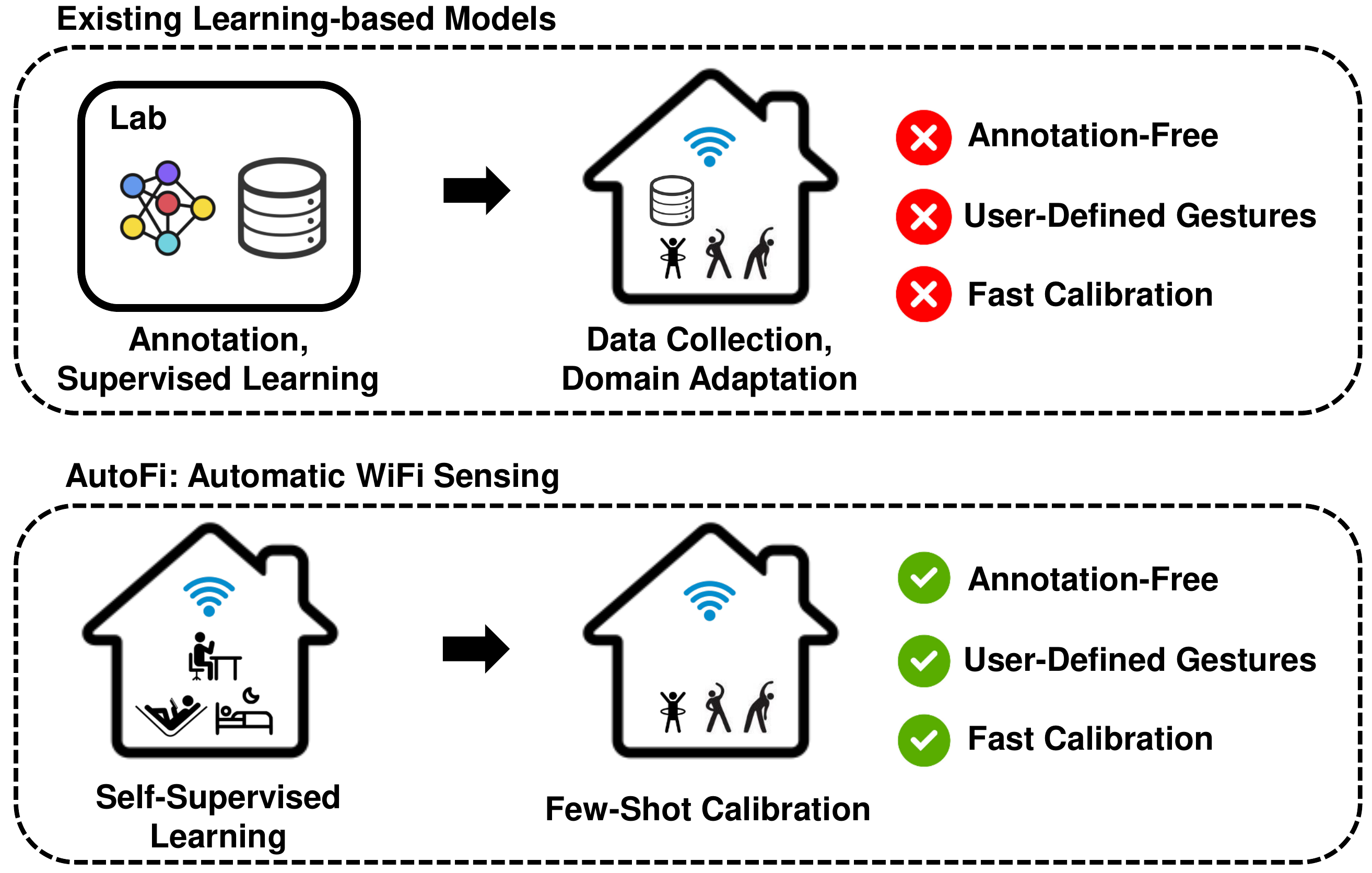}
	\caption{The illustration of the proposed AutoFi framework. Compared to the existing learning-based models, the AutoFi does not require tremendous data collection and annotation in the lab, supports user-defined gestures as new features, and can be setup swiftly and conveniently at user side.}
	\label{fig:intuition}
\end{figure}

% 目前流派：模型 / 机器学习
WiFi sensing methods can be categorized into model-based methods and learning-based methods that serve for different applications. Model-based methods formulate the WiFi signals and its environment by physical models, such as the Fresnel zone~\cite{wang2016human}. For periodic human motions or simple activities such as respiration and falling down~\cite{wang2016rt, hu2021defall}, model-based methods are accurate and robust to environmental variations. However, it is hard to build physical models for complicated activities or compound motions. To deal with it, learning-based models are developed as deep learning models show stronger capacity of extracting and modeling CSI patterns of complex gestures~\cite{zhang2018crosssense}. Nevertheless, the performance and generalization ability of data-driven models depend on the scale and variety of training samples, since the data collection and annotation process is usually time-consuming and labor-intensive. Model-based methods have achieved remarkable performance and robustness, so we mainly study the gap between current learning-based methods and real-world challenging and practical applications.

% 机器学习流派的主要问题是对于环境的依赖，很多方法用DA来解决，但是DA依然需要无监督地收集大量相对均衡的高质量样本，这即使是在客户端也是不可行的，是一个太强的假设。
Generally speaking, learning-based methods rely on statistical or deep learning models that map the CSI data to the label space in terms of specific tasks, such as the identity for human identification or gesture category for gesture recognition~\cite{zou2018deepsense}. It is noteworthy that the success of deep learning models for visual recognition is dependent on the scale of the dataset, e.g. the large-scale ImageNet~\cite{deng2009imagenet}, but such scale of dataset does not exist in WiFi sensing. The reason lies in the difficulty of collecting CSI samples by thousands of volunteers under thousands of circumstances. Recent work contributes to a bigger dataset, such as Widar~\cite{zhang2021widar3}, but its scale is still below the ImageNet. Without sufficient data, learning-based models may fail in a new environment. Then many works commence to explore domain adaptation to deal with cross-environment problems, such as EI system~\cite{jiang2018towards} and WiADG~\cite{zou2018robust}. These works are based on domain adaptation methods that adapt the model to a new environment by minimizing the distribution discrepancy of the feature spaces between training and testing scenarios, which significantly improves the performance in the new environment~\cite{yang2020mobileda}. However, to enable domain adaptation methods, we need to collect a great number of high-quality CSI samples in the new environment, though in an unlabeled manner, but the data should be large-scale and balanced to all categories. Such assumption is naturally hard to achieve for real-world applications where users still need to do laborious data collection. 

% 针对这个问题，我们能否让WiFi模型在新的环境里，利用随机采集的样本，不限定类别和均衡性，训练得到一个好的机器学习模型，并且只用很少量的样本驱动它。
To bridge the gap between learning-based models and realistic WiFi sensing, we study how deep models can work in an automatic data-efficient manner in this paper. In realistic WiFi scenarios, two kinds of data are accessible. Firstly, CSI samples of human daily activities can be obtained without the activity labels and the segmentation of activities in Widar~\cite{zhang2021widar3}. This can be simply achieved by setting a variation threshold of CSI streams, which offers massive unlabeled low-quality CSI samples. Secondly, a few number of labeled data can be collected with the cooperation from user for calibration purpose, which is similar to the existing mobile phone security system setup of face and fingerprint recognition. If these easily-collected data can be leveraged for learning-based models, then it is not necessary to train a model in advance and conduct the domain adaptation process. The whole model learning process is therefore automatic without manual data collection and annotations, and the system can be initiated by users easily.

% 这篇文章中，AutoFi就是来解决这个问题的。
To this end, we propose an annotation-efficient WiFi Sensing system, namely AutoFi, which learns new environmental settings in a self-driven fashion. It is an automatic WiFi representation learning framework that helps achieve automatic WiFi human sensing with very few manual annotations As shown in Figure \ref{fig:intuition}, after deploying the AutoFi in a new environment, AutoFi firstly collects randomly-segmented and randomly-distributed CSI samples for any human actions. These samples could be persons passing by or various daily activities that are easy to acquire. Then, the self-supervised learning module enables the AutoFi to learn CSI patterns in an unsupervised manner, i.e., without the engagement of any labels. After self-supervised learning, the model has been initiated well with new environments learned. Then we can conduct few-shot learning by calibrating several high-quality samples from users. It is worth noting that the task and the gesture categories can be totally customized by users, no matter whether the new defined gestures have been seen or not. It is the first work that achieves cross-task transfer in WiFi sensing. The AutoFi learns how to extract robust features from environmental CSI samples, and contributes to customized functions. Extensive experiments are conducted in the real world and public datasets to demonstrate the effectiveness of our method. 

% 主要贡献
The contributions are summarized as follows:
\begin{itemize}
	\item We analyze the main gaps between learning-based methods and practical WiFi sensing, and propose the AutoFi to deal with it.
	\item In AutoFi, we propose a novel self-supervised learning framework based on prevailing contrastive learning and mutual information, and further enhance its transferability by developing a novel geometric structural loss, which helps the AutoFi to enable various downstream tasks.
	\item The AutoFi achieves the cross-task transfer for WiFi sensing. To the best of our knowledge, it is the first work that achieves automatic WiFi sensing in new environments without any prior data collection.
	\item The AutoFi system is implemented in the real world to validate its robustness. We also simulate the AutoFi using public datasets, e.g., Widar and UT-HAR, and the results are also superior to existing domain adaptive systems.
\end{itemize}

\section{Related Works}
\subsection{WiFi-based Passive Human Sensing}
Recently, WiFi-based passive radar is appealing in smart homes due to its low cost and high granularity. Compared to visual sensing~\cite{vedadi2017automatic}, WiFi sensing is privacy-preserving and illumination-robust. WiFi sensing relies on channel state information that is extracted from specific WiFi chips, such as Intel 5300 NIC~\cite{halperin2011tool} and Atheros NIC~\cite{xie2015precise}. The number of subcarriers and antennas determines the resolution of the CSI data. The Intel 5300 NIC tool can extract 30 subcarriers of CSI from each pair of antennas with a 20Mhz bandwidth, while the Atheros tool can take out 114 subcarriers of CSI with 40Mhz. The CSI data records the surrounding objects or motions that affect the multi-path propagation of wireless signals. This process can be depicted by some physical models, such as Fresnel zone~\cite{wang2016human}. Relying on model analytics and signal processing, WiFi passive radar achieves high performance on detecting periodic motions and specific human activities. The signal tendency index (STI) is developed to identify the occupancy situation~\cite{zou2017freedetector}. Want et al. propose a respiration detection system and investigates the effect of user location and orientation~\cite{wang2016human}, which is very useful in healthcare. Currently, WiFi sensing has widespread applications including occupancy estimation~\cite{zou2017freecount,zou2018device,zou2017freedetector}, activity recognition~\cite{zou2017multiple,zou2017poster,wang2021multimodal,yang2018carefi,yang2018fine,yang2021deep,yang2022metafi}, gesture recognition~\cite{yang2019learning,zou2018joint,zou2018robust}, human identification~\cite{zou2018identification,wang2022caution,zhang2020gate}, human pose estimation~\cite{yang2022metafi} and vital sign detection~\cite{wang2016human,hu2022resfi}.

\subsection{Learning-based Methods for WiFi Sensing}
However, for more complex human gestures or even customized activities by users, machine learning models contribute to better capacity to recognize them. Wang et al. firstly propose a human activity recognition system by statistical features (e.g. mean and peak) and traditional classifiers~\cite{wang2015understanding}. Then the E-eyes system is developed to achieve better performance by dividing human activities into in-place and dynamic ones~\cite{wang2014eyes}. The FreeCount system leverages a feature selection scheme based on information theory to conduct people counting~\cite{FreeCount}. These early-stage works show good performance on normal activities such as walking and sitting, but they cannot identify fine-grained subtle gestures. To enhance the model capacity for these gesture recognition, deep learning models are introduced. Zou et al. propose the DeepSense that learns spatial-temporal features based on the convolutional neural network (CNN) and recurrent neural network~\cite{zou2018deepsense}. Yang et al. propose the EfficientFi that realizes the large-scale WiFi sensing models by learning-based CSI compression~\cite{yang2022efficientfi}. SecureSense is proposed to deal with the adversarial attacks by learning prediction consistency~\cite{yang2022securesense}. Chen et al. propose a bi-directional LSTM for activity recognition~\cite{chen2018wifi}. These machine learning and deep learning methods show great performance in a single environment, but cannot generalize well to a new environment. To address this issue, adversarial domain adaptation methods transfer knowledge from a source domain to a new target domain using only unlabeled examples~\cite{zou2018robust}. Then domain adaptation~\cite{yang2020mobileda} is a prevailing method for cross-environment WiFi sensing applications, such as TransferSense~\cite{bu2021transfersense}. Nevertheless, it is noted that we still need high-quality CSI samples that have same categories, balanced label distribution and well-segmented actions in the unlabeled target domain~\cite{zhang2021privacy}, which requires users to engage and thus is still cumbersome. Another solution is to generate target-like samples by the generative adversarial network, but this also demands a number of high-quality data~\cite{wang2021multimodal}. Our proposed AutoFi deals with this problem by learning randomly-segmented and randomly-distributed samples for downstream tasks, and hence it achieves automatic learning models for WiFi sensing in the real world.

\subsection{Self-Supervised Learning and Few-shot Learning}
As the AutoFi consists of two phases based on self-supervised learning and few-shot learning, we also review some recent progress on these perspectives. Self-supervised learning is a promising method to learn feature representations in an unsupervised manner~\cite{jing2020self}. Previous self-supervised methods are designed for unsupervised visual feature learning, and they mainly rely on designing handcrafted auxiliary tasks, such as context prediction~\cite{doersch2015unsupervised} and rotation prediction~\cite{gidaris2018unsupervised}. They achieve good performance but the handcrafted tasks limit the generalization ability of models. Then constrastive methods come into existence~\cite{jaiswal2020survey}, which learns features from multiple views of samples via metric learning. SimCLR proposes to minimize the cosine similarity between views of same samples and maximize the similarity between those of different samples~\cite{chen2020simple}. Then the BYOL~\cite{grill2020bootstrap} firstly abandons the negative samples and adopt asymmetric architecture to mitigate the collapsed solution. Maximizing mutual information for representation learning is also prevailing, such as Deep InfoMax~\cite{hjelm2018learning} and TWIST~\cite{wang2021self}. Though self-supervised learning helps generate a discriminative feature space, it does not contain any supervision tasks. To enable real-world applications, we further consider a data-efficient learning scheme: few-shot learning. Few-shot learning aims to conduct classification or regression by learning only several samples, or even one sample (i.e. one-shot learning)~\cite{sung2018learning}. It is highly related to metric learning that is widely applied to face recognition~\cite{schroff2015facenet}, where triplet loss is utilized to cluster the samples from the same category and separate the samples from different categories. Yang et al. propose to leverage few-shot learning for WiFi-based gesture recognition. However, in few-shot learning in a new environment, we still need to initialize the model parameters using labeled training data collected in another environment, and this may lead to a domain shift that hinders the model performance. In the AutoFi, we enable the model to learn the environment by itself, and then utilize few-shot learning for gesture recognition.

\begin{figure*}[t]
	\centering
	\includegraphics[width=1.0\linewidth]{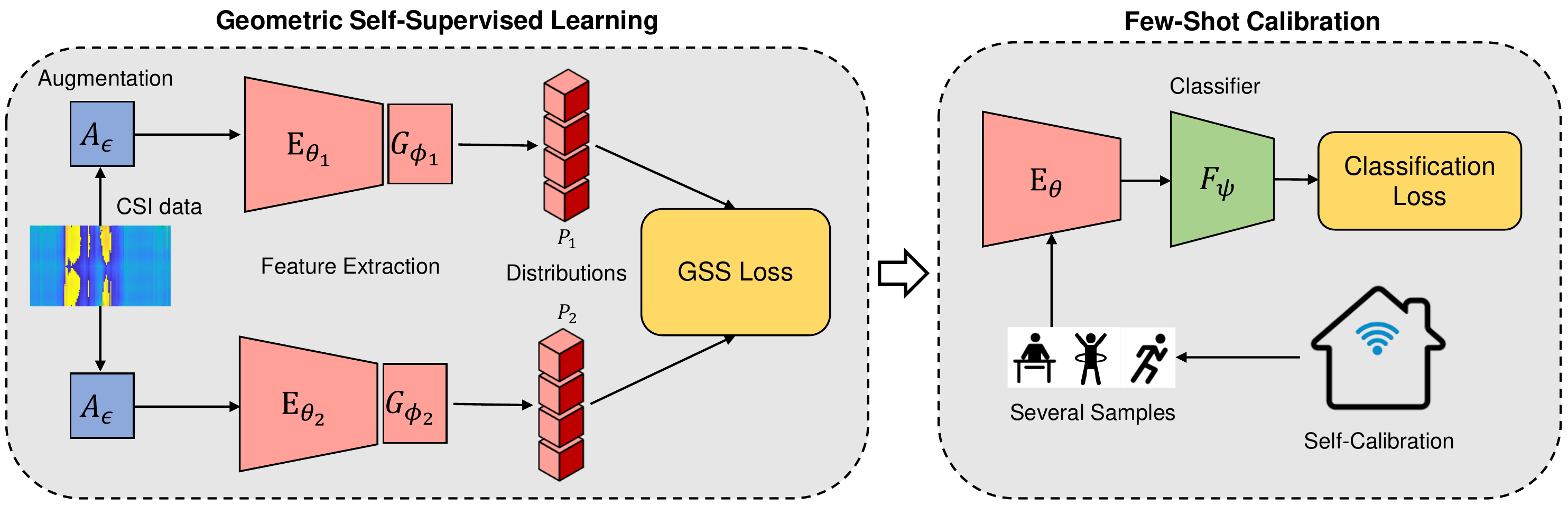}
	\caption{The illustration of the proposed AutoFi learning method. The AutoFi consists of a geometric self-supervised learning module that learns CSI features from randomly-collected CSI samples, and a few-shot calibration module that enables users to easily enable the recognition services. The feature extractor $E_\theta$ in the few-shot calibration is initialized by the self-supervised module.}
	\label{fig:framework}
\end{figure*}

\section{Method}
\subsection{Overview}
The objective of the AutoFi design is to enable learning-based WiFi sensing by minimizing manual efforts. As shown in Figure \ref{fig:framework}, the AutoFi is composed of two modules: a geometric self-supervised learning module and a few-shot calibration module. In the self-supervised learning module, the randomly-collected CSI data is processed by an augmentation $A_{\epsilon}$ to generate two random views, and these two views are fed into the feature extractors $E_{\theta_1},E_{\theta_2}$ and the non-linear functions $G_{\phi_1},G_{\phi_2}$ to produce two distributions. The geometric self-supervised (GSS) loss enforces these two prediction distributions to be consistent, which does not require any annotations. Then the well-trained feature extractors $E_{\theta_1},E_{\theta_2}$ can be transferred to the few-shot calibration module. Users only need to calibrate some gestures for several times to enable the recognition system, which allows users to define customized gestures or tasks. For the few-shot training, we use the prototypical network as the backbone~\cite{snell2017prototypical}.

\subsection{Geometric Self-Supervised Learning Module}
The geometric self-supervised (GSS) learning module aims to learn CSI representations in an unsupervised manner. Prevailing self-supervised learning methods employ handcrafted auxiliary tasks or contrastive learning~\cite{jing2020self}. In our scenarios, the downstream tasks can be quite different from the training samples that are randomly collected, and thus requires better transferability and generalization ability, which motivates us to design the GSS based on contrastive learning due to its stronger generalization capability~\cite{grill2020bootstrap}. The GSS modules consists of an augmentation module $A_{\epsilon}$ with a hyper-parameter $\epsilon$, the feature extractors $E_{\theta_1},E_{\theta_2}$ parameterized by $\theta_1,\theta_2$, respectively, and the non-linear functions $G_{\phi_1},G_{\phi_2}$ parameterized by $\phi_1,\phi_2$, respectively. The feature extractors are normally CNNs and the non-linear functions are just multilayer perceptrons (MLPs). The input data is the randomly-collected unlabeled CSI samples $\{x^i\}_{i=1}^{N}$. Each CSI sample is a matrix such that $x^i \in \mathbb{R}^{S\times T}$ where $S$ denotes the number of subcarriers and $T$ denotes the time duration.

\textbf{Multi-view Generation.} Firstly, we input the samples to the augmentation module $A_{\epsilon}$. The augmentation module aims to generate two views for the self-supervised learning. The two views should be meaningful, but randomly augmented, such as the random cropping for images. For CSI data, previous research shows that the noises on subcarriers can be modeld as Gaussian noise~\cite{yang2018carefi}. Hence, without break the intrinsic information of the CSI data, we augment the input sample by adding a Gaussian noise $\zeta \sim \mathcal{N}(\mu,\sigma^2)$:
\begin{equation}
	A_{\epsilon}(x^i) = x^i + \epsilon \zeta,
\end{equation}
where $\epsilon$ is the weight of the noise. We can generate two views $x^i_1,x^i_2$ by $A_{\epsilon}(x^i)$.

The next step is to extract features by $E_{\theta_1}$. Here we just leverage a series of convolutional layers for $E_{\theta_1}$ as successfully used in many previous works~\cite{zou2018deepsense}. Then the feature embeddings are generated, but this feature space is what we aim to do classification in the few-shot learning. For self-supervised learning, we need to separate the feature space by a non-linear function $G_{\phi_1}$. The bottleneck layer $G_{\phi_1}$ ensures that the self-supervised learning will not affect the feature learning, as discovered in~\cite{grill2020bootstrap}. After $E_{\theta_1}$ and $G_{\phi_1}$, the feature distributions of the first view are calculated by
\begin{equation}
	P(x^i_1)=G_{\phi_1}(E_{\theta_1}(A_{\epsilon}(x^i_1))).
\end{equation}
The second view is processed by $E_{\theta_2}$ and $G_{\phi_2}$ in the same way. In this fashion, $P(x^i_1)$ and $P(x^i_1)$ are obtained.

\textbf{Probability Consistency.} How to design the unsupervised loss is the key of the GSS module. We propose a novel learning objective that firstly incorporates geometric structures for unsupervised learning, which can benefit the downstream few-shot task. In contrastive learning, the normal objective is to force the predictions of different views to be consistent. To this end, the probability consistency loss is formulated as
\begin{equation}
	\mathcal{L}_p=\frac{1}{2B}\sum_{i=1}^B (D_{KL}(P_1^i||P_2^i)+D_{KL}(P_2^i||P_1^i)),
\end{equation}
where $D_{KL}(\cdot||\cdot)$ denotes the Kullback–Leibler divergence of the two distributions. Since the KL divergence is an asymmetric measure of distributions, we use dual forms to make it symmetric. By the consistency loss, the model learns to perform consistently on two views in terms of the prediction probabilities.

\textbf{Mutual Information.} In our scenario, we require the feature extractor to have the transferability for downstream tasks. To this end, we aim to maximize the mutual information between CSI samples and the feature space for better transferability. From the information theory, the mutual information between the prediction distributions and the input space should be maximized. The mutual information between a random variable $X$ and its predicted label $Y$ is formulated by
\begin{equation}
	I(X,Y)=H(Y)-H(Y|X),
\end{equation}
where $H(\cdot)$ is the information entropy. Increasing $H(Y)$ drives the model to predict uniform distributions among classes, while decreasing $H(Y|X)$ drives the model confidence of its predictions. However, the mutual information cannot be calculated directly, and therefore we aim to maximize its approximation by
\begin{equation}
	\mathcal{L}_m=h(\mathbb{E}_{x^i\in \mathcal{B}}P^i)+ \mathbb{E}_{x^i\in \mathcal{B}}h(P^i),
\end{equation}
where $\mathcal{B}$ is a batch of samples and $h(p)=-\sum_i p_i \log p_i$ is the conditional entropy. The $\mathcal{L}_m$ operates on both $P_1$ and $P_2$ for all samples. The mutual information loss is widely used in semi-supervised learning and domain adaptation~\cite{liang2020we}.

{\color{black}
\textbf{Geometric Consistency.} For our system, apart from learning discriminative features from unlabeled CSI samples, we further require the AutoFi to empower recognition capacity via few-shot learning. Nevertheless, former self-supervised learning may not be tailored for this purpose. They mostly rely on the probability consistency and information maximization that enable a discriminative feature space, but do not consider the downstream few-shot tasks. To deal with this problem, we propose a novel geometric loss in the GSS module. The rational behind this stems from the feature space of few-shot learning. The few-shot learning is highly related to metric learning and prototypical networks~\cite{snell2017prototypical,sung2018learning} which leverage the cluster of each category and their geometric relationship. With tight clusters and meaningful geometry, the test sample can be predicted by retrieving the category of the most similar sample or applying k-nearest neighbors strategy in the feature space. In our scenarios, traditional self-supervised learning frameworks fail to capture geometry while classic few-shot learning frameworks cannot work well due to the lack of labels. To utilize the geometry among unlabeled samples, we propose a geometric structural loss that forces the geometry of two batches of views to be consistent. The geometry of a batch of samples can be generated by the relationship of neighbors. For a sample $x^i$ with distribution $P^i$, its geometric embedding $Q^i$ can be formulated as
\begin{equation}
	q_{i|j}=\frac{K(P^i,P^j)}{\sum_{m=1,m\neq j}^B K(P^m,P^j)},
\end{equation}
where $q_{i|j}$ denotes the $j$-th position of $Q^i$, and $K(\cdot,\cdot)$ is a similarity function. Here we choose the cosine similarity as
\begin{equation}
	K(\mathbf{a},\mathbf{b}) = \frac{1}{2}(\frac{\mathbf{a}^T \mathbf{b}}{\|\mathbf{a}\|_2 \|\mathbf{b}\|_2}+1).
\end{equation}
Note that the geometric embedding $Q^i$ represents the relationship between $x^i$ and all neighbors in the feature space. Then we train the model to generate a consistent geometry on two views by applying KL divergence:
\begin{equation}
	\mathcal{L}_g=D_{KL}(Q_1^i||Q_2^i).
\end{equation}
The geometric structural loss helps model learn geometry of CSI samples and further learn the feature space in terms of metrics. In this manner, the GSS module can enhance the subsequent few-shot learning module.
}

The total objective of the loss is defined as
\begin{equation}
	\mathcal{L} = \mathcal{L}_p + \lambda \mathcal{L}_m + \gamma \mathcal{L}_g,
\end{equation}
where $\lambda$ and $\gamma$ are two hyper-parameters that balance multiple objectives for better convergence. In self-supervised learning, as long as they have similar magnitudes, the convergence can be achieved easily.

\subsection{Few-shot Calibration Module}
After the GSS module, we transfer the feature extractors $E_{\theta_1},E_{\theta_2}$ to the few-shot calibration (FSC) module, and re-use it to train a classifier for few-shot learning. Note that the two feature extractors are very similar, so either one can be used in FSC, denoted as $E_\theta$. Users only need to collect several samples to setup the AutoFi. The labeled samples are denoted as $\{x^i,y^i\}_{i=1}^M$ where $M$ is the number of labeled samples. The feature embedding can be obtained by feeding samples into the feature extractor $E_\theta$, and a classifier $F_\psi$ maps the feature to its labels. In few-shot calibration, we firstly minimize the standard cross-entropy loss:
\begin{equation}
	\mathcal{L}_{c}=-\mathbb{E}_{(x,y)}\sum_k \big[ \mathbb{I}[y=k] \log \big(F_\psi(E_\theta(x^i))) \big],
\end{equation}
where $\mathbb{I}[y=k]$ means a 0-1 function that outputs 1 for the correct category $k$. Then to better cluster the same-class samples, we calculate the prototypes of each class as $c_k$, and draw the same-class samples together by minimizing the log-probability
\begin{equation}
	\mathcal{L}_{f}=-\log p_{\theta,\psi}(y=k|x),
\end{equation}
where $p_{\theta,\psi}(y=k|x)$ is constructed by the distance between the sample $x^i$ and its correct class center, formulated as
\begin{equation}
	p_{\theta,\psi}(y=k|x^i)=\frac{\exp (-d(F_\psi(E_\theta(x^i)),c_k))}{\sum_{k'}\exp (-d(F_\psi(E_\theta(x^i)),c_{k'}))},
\end{equation}
where $k'$ denotes all categories. Note that the gesture or activity category, and even the recognition task can be customized by users. The few-shot calibration is a normal few-shot learning scheme motivated by prototypical network~\cite{snell2017prototypical}. Whereas, after the feature extractor learns the randomly-collected samples in the GSS, it is found that the convergence of the FSC module can be easily achieved and the performance is boosted. In this manner, the AutoFi can quickly adapt to any environment automatically, and users input can enable the AutoFi to perform many downstream tasks without cumbersome data collection and model training. The whole algorithm is illustrated in Algorithm~\ref{algo:autofi}.

\begin{algorithm}[t]
	\small
	\LinesNumbered
	\SetAlgoLined
	\SetAlgoLongEnd
	\DontPrintSemicolon
	\SetKwInput{KwModule}{Module}
	\SetKw{KwBegin}{BEGIN:}
	\SetKw{KwEnd}{END.}
	
	\small
	\caption{Automatic WiFi Sensing Setup\label{algo:autofi}}
	
	\textbf{Step 1: Train the GSS module}\;
	\KwModule{
		the feature extractors $E_{\theta_1},E_{\theta_2}$, the non-linear functions $G_{\phi_1},G_{\phi_2}$
	}
	\KwIn{unlabeled CSI data $\{x^i\}_{i=1}^{N}$}
	
	\KwBegin{}\;
	\While(){epoch $<$ total epoch}{
		
		Augment samples by $A_{\epsilon}(x^i) = x^i + \epsilon \zeta$\;

		Obtain feature probabilities of views via $P(x^i_1)=G_{\phi_1}(E_{\theta_1}(A_{\epsilon}(x^i_1)))$\;

		Update $\theta_1,\theta_2,\phi_1,\phi_2$ by minimizing $\mathcal{L}_p + \lambda \mathcal{L}_m + \gamma \mathcal{L}_g$
	}
	
	\KwOut{the model parameters $\theta_1,\theta_2$.}
	\textbf{Step 2: Train the FSC module}\;
	\KwModule{the classifier $F_\psi$}
	\KwIn{a few labeled samples $\{x^i,y^i\}_{i=1}^M$}
	Initialize $E_{\theta}$ by either $\theta_1$ or $\theta_2$.
	\While(){epoch $<$ total epoch}{		
		Update $\theta,F_\psi$ by minimizing $\mathcal{L}_c + \mathcal{L}_f$
	}
	\KwOut{the model parameters $\theta,\psi$.}
	\KwEnd
\end{algorithm}

\section{Experiments}
\subsection{Setup}
\textbf{Evaluation Scenarios and Criterion.} We evaluate the AutoFi on different WiFi platforms and CSI data. Firstly, the AutoFi is implemented on a real-world IoT system for evaluation, demonstrating the main novelty of the AutoFi - to learn the environment by self-supervised learning and perform downstream tasks by few shots. The real-time system is based on Atheros CSI tool and fine-grained CSI data~\cite{xie2015precise}. Then, we evaluate the effectiveness of the AutoFi using UT-HAR dataset, which leverages Intel 5300 NIC with a sampled number (30) of CSI subcarriers~\cite{yousefi2017survey}. The third experiments are conducted on a large dataset, Widar~\cite{zhang2021widar3}. Due to the different collection scenario, it is used to demonstrate that the AutoFi can support new types of gestures after self-supervised learning. The criterion is the top-1 accuracy across all test samples.

\textbf{Implementation Details.} Here we introduce the details of the AutoFi, and the experimental settings are introduced in the following subsections. The two modules of the AutoFi are implemented by Pytorch. The network structures are shown in Table~\ref{table:network}. The SGD optimizer is utilized with a learning rate of 0.01 and a momentum of 0.9. The epoch of training GSS module is 300 and the FSC is trained for 100 epochs. The batch size is set to 128 in order that the GSS module can capture the geometry among samples. For all the experiments, we set the hyper-parameter $\lambda=1$ and $\gamma=1000$, which aims to keep the magnitudes of multiple losses similar. 

\textbf{Baselines.} As our method mainly deals with the few-shot learning scenario, we compare our method with recent state-of-the-art few-shot recognition methods based on CSI, including the CSI-GDAM~\cite{zhang2021csi}, the ReWiS~\cite{bahadori2022rewis}, and the classic prototypical network~\cite{snell2017prototypical} that is the baseline method. The CSI-GRAM utilizes the graph neural network and attention scheme to enhance few-shot learning, while the ReWiS proposes SVD data processing and applies the prototypical network.

\begin{table}[]
	\centering
	\begin{tabular}{c|c|c}
		\toprule \midrule
		Layer Index & Feature Extractor $E_\theta$         & Classifier $F_\psi$   \\ \midrule
		input           & \multicolumn{2}{c}{CSI data: 3 $\times$ 114 $\times$ 500}                         \\ \midrule
		1           & Conv 32$\times$(15,23), stride 9, ReLU            & 128 dense \\ \midrule
		2           & Conv 32$\times$(3,7), stride 1, ReLU     &   6 dense, softmax                \\ \midrule
		3           & Max-pool (1,2), stride (1,2)   &                   \\ \midrule
		4           & Conv 64$\times$(3,7), stride 1, ReLU    &                   \\ \midrule
		5           & Conv 96$\times$(3,7), stride 1, ReLU  &        \\ \midrule
		6           & Max-pool (1,2), stride (1,2)   & \\ \midrule \bottomrule
	\end{tabular}
	\caption{The network architecture used in the AutoFi experiments. For Conv A$\times$(H,W), A denotes the channel number, and (H,W) represents the height and width of the operation kernel. This applies to all Convolution (Conv) and Max-pooling (Max-pool) layers.}\label{table:network}
\end{table}

\begin{figure}[t]
	\centering
	\includegraphics[width=1.0\linewidth]{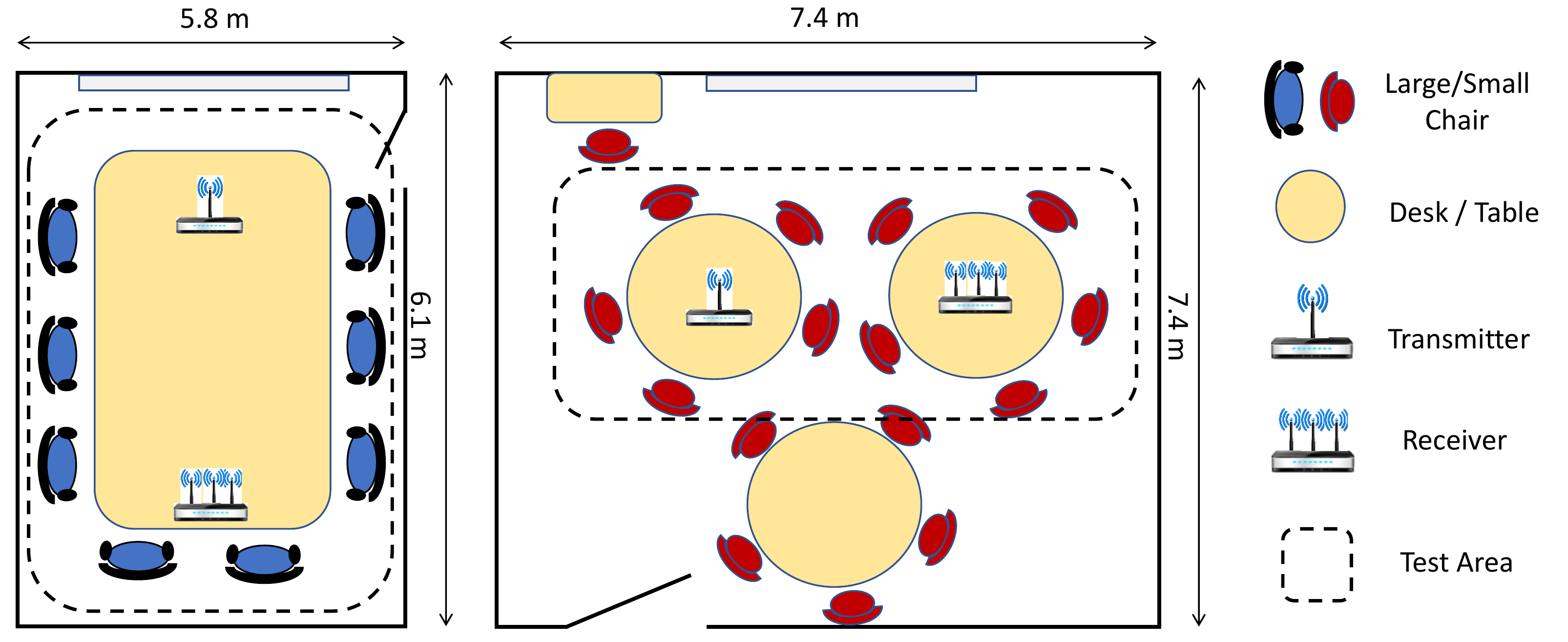}
	\caption{The layouts of the AutoFi experiments.}
	\label{fig:layout}
\end{figure}

\begin{table*}[t]
	\caption{Accuracy (\%) of the AutoFi in the real-world experiments.}\label{tab:exp-real-world}
	\centering
	\begin{tabular}{l|cccc|cccc}
		\toprule
		\multirow{2}{*}{Method} & \multicolumn{4}{c|}{Gesture Recognition}         & \multicolumn{4}{c}{Human Identification}         \\
		& 1-shot         & 2-shots        & 3-shots    &   Avg    & 1-shot         & 2-shots        & 3-shots      &   Avg     \\ \midrule
		Prototypical Network~\cite{snell2017prototypical}   & 77.65          & 82.76          & 85.42   &  81.94       & 67.62          & 74.29          & 78.51   &  73.47       \\
		ReWiS~\cite{bahadori2022rewis}                  & 77.63          & 81.38          & 86.47    &   81.83      & 68.13          & 77.65          & 77.85    &   74.54      \\
		CSI-GDAM~\cite{zhang2021csi}                & 81.64          & 83.92          & 85.37     &  83.64      & 69.38          & 77.72          & 79.67    & 75.59      \\
		AutoFi                  & \textbf{83.31} & \textbf{87.46} & \textbf{89.71} & \textbf{86.83} & \textbf{72.86} & \textbf{82.65} & \textbf{83.33} & \textbf{79.61} \\ \bottomrule
	\end{tabular}
\end{table*}

\begin{table*}[t]
	\caption{Accuracy (\%) of the AutoFi on human identification in three situations.}\label{tab:human-id}
	\centering
	\begin{tabular}{l|ccc|ccc|ccc}
		\toprule
		\multirow{2}{*}{Method} & \multicolumn{3}{c|}{(a) Jacket}                  & \multicolumn{3}{c|}{(b) Backpacks}               & \multicolumn{3}{c}{(c) Dynamics} \\
		& 1-shot         & 2-shots        & 3-shots        & 1-shot         & 2-shots        & 3-shots        & 1-shot   & 2-shots   & 3-shots   \\ \midrule
		Prototypical Network~\cite{snell2017prototypical}   & 37.14          & 50.00          & 61.56          & 39.52          & 45.24          & 55.78          & 45.24    & 49.05     & 54.40     \\
		AutoFi                  & \textbf{64.39} & \textbf{62.34} & \textbf{66.00} & \textbf{50.90} & \textbf{58.10} & \textbf{59.86} & \textbf{61.22}    & \textbf{62.86}     & \textbf{63.81}     \\ \bottomrule
	\end{tabular}
\end{table*}

\subsection{Real-world System Evaluation}
{\color{black}
\textbf{System Setup.} To demonstrate the effectiveness of our AutoFi, we implement our system in the real world. The AutoFi system consists of two TPLink-N750 routers that serve as the transmitter and receiver. They are set to operate on 5Ghz with a bandwidth of 40MHz. Leveraging Atheros CSI tool~\cite{xie2015precise} and real-time IoT platform~\cite{yang2018device}, we extract the 114 subcarriers of CSI data for each pair of antennas. The receiver is equipped with 3 antennas while the transmitter is equipped with 1 antenna. The sampling rate is 100Hz and each CSI sample is captured for 5 seconds with the size of $3\times 114\times 500$. Only CSI amplitudes are used since the phase information is not stable for the Atheros tool. As shown in Figure~\ref{fig:layout}, we evaluate the AutoFi in two different environments. The first environment only has one table and all chairs are surrounded, while the second one has a more complicated layout with four tables and many chairs. We set a threshold $\tau=20$ to capture CSI samples randomly. As long as the CSI amplitude is greater than $\tau$, the system starts to record the CSI data for 5s. In this way, we leave the AutoFi system alone for automatic data collection, and we obtain more than 5000 samples without any human labor for the self-supervised learning. This automatic data collection process took about half a day. Then we collect very few labeled CSI samples to conduct few-shot calibration, which can be easily achieved up to several minutes in the real world as only 1-3 samples are required for one gesture. The downstream tasks are the gesture recognition in the first environment, and the human gait recognition in the second environment. The test samples are collected anywhere within the regions, and they are annotated only for to serve as ground truth for performance. For gesture recognition, there are 8 types of gestures including up \& down, left \& right, pull \& push, clap, fist, circling, throw, and zoom, with 120 samples from each category for testing. For human identification, 14 volunteers are engaged with 20 samples from each category for testing. The volunteer walks though the line of sight of the two routers either with a jacket or a backpack, which makes the task challenging. Two experiments are independently conducted, and there exist some environmental dynamics as some staff are working around. No data preprocessing techniques are utilized for model training.
}

\textbf{Results.} According to different shots of few-shot calibration, we summarize the overall results in the Table~\ref{tab:exp-real-world}. It is seen that the AutoFi achieves 83.31\%, 87.46\% and 89.71\% accuracy on gesture recognition task with 1-shot, 2-shots, and 3-shots learning, respectively, outperforming the baseline method by 4-6\%. For the human identification task, more categories and the heterogeneity of gaits lead to more challenges. The overall accuracy is worse than the accuracy on the gesture recognition task. The AutoFi still achieves the state-of-the-art performance when it is compared to the ReWiS and CSI-GDAM. It is seen that the ReWiS only slightly outperforms the prototypical network, while the CSI-GDAM attains a stable improvement. In summary, the GSS module of the AutoFi learns the environmental dependency, and thus promotes the subsequent few-shot learning by the prototypical network. The results demonstrate that the AutoFi can learn randomly-collected samples by itself, and transfer the knowledge to distinct downstream tasks.

\textbf{Feature Transferability.} For human identification, we have three testing scenarios: (a) subjects wearing jacket, (b) subjects wearing backpacks, and (c) subjects wearing jacket and backpacks with enhanced environmental dynamics. We let the AutoFi only incorporate few-shot samples from a single scenarios and test it on all scenarios, which verifies the transferability ability of the features. We compare it with the single prototypical network in the Table~\ref{tab:human-id}. It is noted that our proposed AutoFi achieves significant improvements across all tasks. Especially, it improves the baseline method on one shot learning for subjects in jacket by 27.25\%. This demonstrates that the features learned by our method have strong transferability. Moreover, it is obvious that the situation of subjects in jacket has the best results for the AutoFi. The reason is that the jacket or backpacks are interference in supervised learning, which may dominate the classifier. The learning-based models are prone to learn these irrelevant features because these may help identification but only for samples, not identity. For example, the backpack may swing as the subject passes by, which helps classification but not human identification. This further shows the importance of feature transferability, and the negative effect of corrupt samples for normal few-shot learning.

\subsection{Evaluation on UT-HAR Dataset}
\textbf{Data Setup.} The UT-HAR~\cite{yousefi2017survey} is a human activity recognition dataset collected by the University of Toronto. There are 7 categories including \textit{lie down, fall, walk, run, sit down, stand up} and \textit{empty}. The sampling rate is 1000Hz that is too large for an input, and the dataset is continuously without segmentation. Therefore, we can simulate our scenario by randomly segmenting the dataset into pieces of CSI for the self-supervised training, and then conduct the few-shot testing. To this end, we segment the data randomly and get 3977 CSI samples. We prepare 10 and 20 labeled samples per category for few-shot calibration, and 70 samples per category for evaluation, which forms the 10-shots and 20-shots activity recognition problem. The size of the input data is $3\times30\times250$. The first layer of the GSS module is slightly modified to match the input size. 

\begin{figure}[t]
	\centering
	\includegraphics[width=0.9\linewidth]{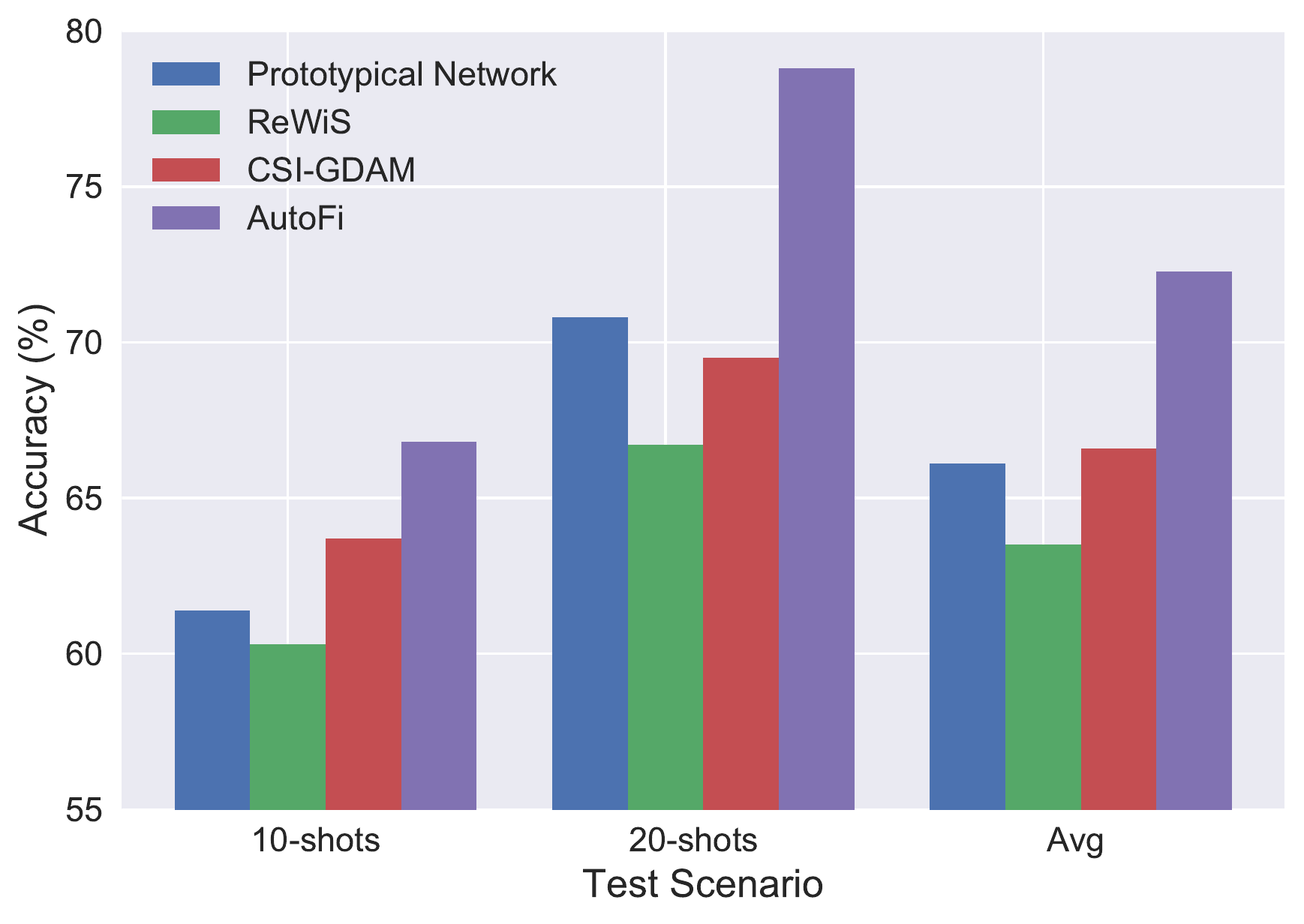}
	\caption{Accuracy (\%) comparison on UT-HAR~\cite{yousefi2017survey} dataset.}
	\label{fig:ut-har}
\end{figure}

\textbf{Results.} The results are shown in the Figure~\ref{fig:ut-har}. The proposed AutoFi achieves the accuracy of 66.8\% and 78.8\% on 10-shots and 20-shots tasks, which demonstrates the effectiveness of our method. Nevertheless, the overall performances are lower than those of the real-world evaluation. The reason are two-folds. Firstly, the UT-HAR dataset is a not well segmented dataset, so there still exists noises for few-shot training samples. Such noise hinders the training significantly. Secondly, the dataset is collected using the Intel 5300 NIC~\cite{halperin2011tool} that only supports 30 subcarriers for each pair of antenna. The resolution is much lower than ours (i.e. 114 subcarriers). It is seen that the low resolution and data noises decrease the performance of few-shot learning.

%\begin{table}[]
%	\centering
%	\caption{Accuracy (\%) of the AutoFi on UT-HAR~\cite{yousefi2017survey} in three situations.}\label{tab:ut-har}
%	\begin{tabular}{l|cc|c}
%		\toprule
%		\multirow{2}{*}{Method} & \multicolumn{2}{c|}{UT-HAR} & \multirow{2}{*}{Avg}  \\ 
%		& 10-shots      & 20-shots      \\ \midrule
%		Prototypical Network~\cite{snell2017prototypical}   & 61.40          & 70.80   &   66.10     \\
%		ReWiS~\cite{bahadori2022rewis}                  & 60.30          & 66.70    &    63.50      \\
%		CSI-GDAM~\cite{zhang2021csi}                &  63.70        & 69.50   & 66.60  \\
%		AutoFi                  & \textbf{66.80} & \textbf{78.80} & \textbf{72.80} \\ \bottomrule
%	\end{tabular}
%\end{table}

\begin{figure}[t]
	\centering
	\includegraphics[width=0.9\linewidth]{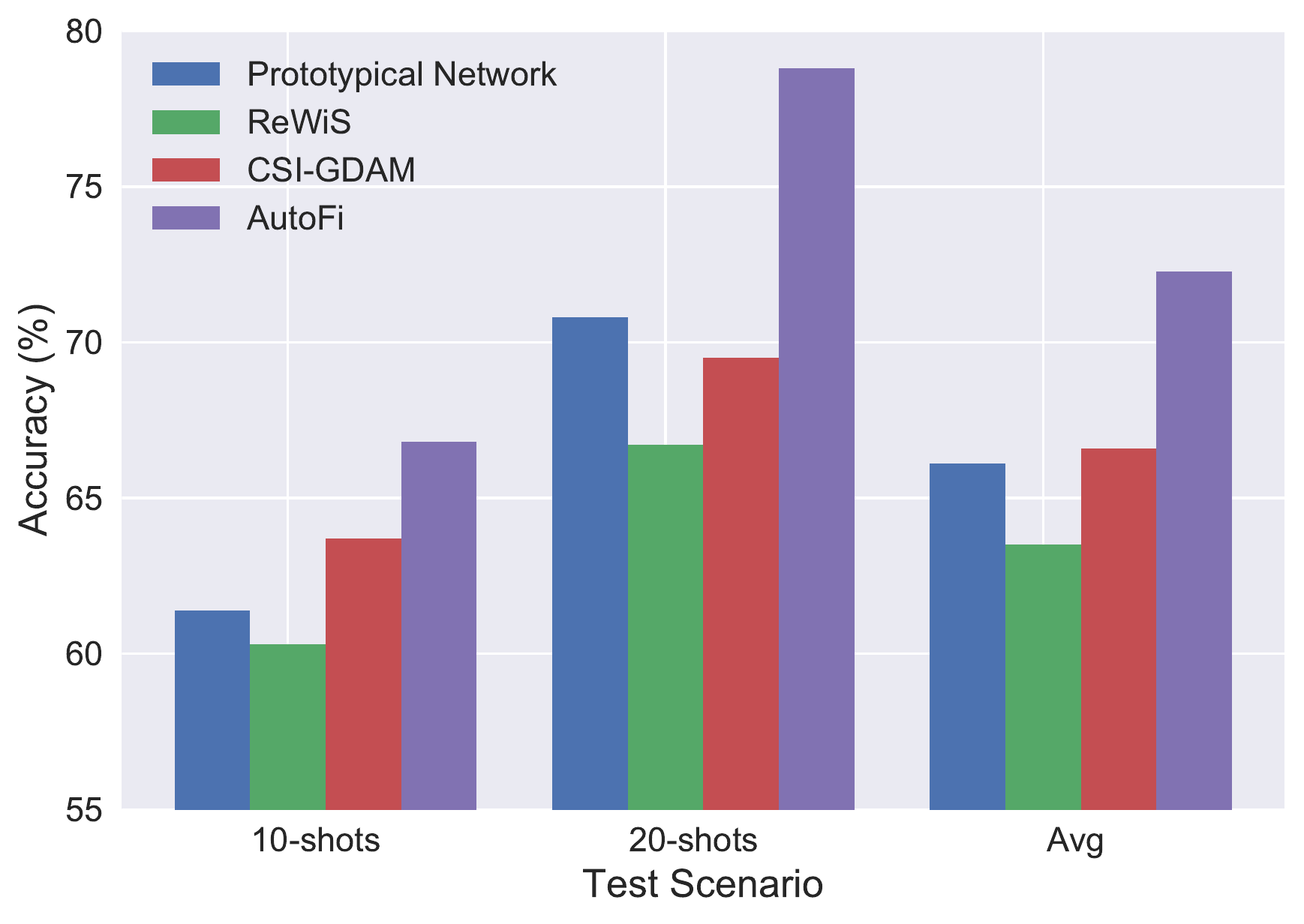}
	\caption{Accuracy (\%) comparison on Widar~\cite{zhang2021widar3} dataset.}
	\label{fig:widar}
\end{figure}

\subsection{Evaluation on Widar Dataset}
\textbf{Data Setup.} Since the UT-HAR has intrinsic noises, we further investigate a large-scale dataset collected by Intel 5300 NIC, the Widar~\cite{zhang2021widar3}. In this dataset, we directly use its transformed data, namely Body-coordinate Velocity Profile (BVP), which eliminates the influence of environment noises. The size of the BVP is $20\times20\times T$, and $T=40$ is the duration. In this experiment, we aim to further demonstrate that \textit{the AutoFi helps increase the feature transferability in terms of new categories} for other data modalities of CSI data. To this end, we use 16 categories of gestures for self-supervised learning, and 6 gestures for few-shot calibration. The first layer of the GSS module is slightly modified to match the input size. 

\textbf{Results.} As shown in the Figure~\ref{fig:widar}, the proposed AutoFi achieves 55.60\% and 63.80\% accuracy for 10-shots and 20-shots recognition tasks, respectively, outperforming the baseline method by 14.40\% and 8.5\%, respectively. It is observed that the ReWiS does not achieve improvement, and the possible reason is that the SVD method may not work for BVP. The overall performance on Widar is worse than that of UT-HAR and our real-world experiments, since the testing data here does not come from one environment, which actually does not conform with our scenario. Nevertheless, we use this dataset to demonstrate that the AutoFi can realize the enlargement of the gestures for the CSI-based gesture recognition system. Even though the training categories for the GSS are not overlapped with the testing categories and the environment varies, the AutoFi can still bring significant improvement for existing methods. 

%\begin{table}[]
%	\centering
%	\caption{Accuracy (\%) of the AutoFi on Widar~\cite{zhang2021widar3} in three situations.}\label{tab:widar}
%	\begin{tabular}{l|cc|c}
%		\toprule
%		\multirow{2}{*}{Method} & \multicolumn{2}{c|}{Widar} & \multirow{2}{*}{Avg}  \\
%		& 10-shots      & 20-shots      \\ \midrule
%		Prototypical Network~\cite{snell2017prototypical}   & 41.20          & 55.30     &   48.25   \\
%		ReWiS~\cite{bahadori2022rewis}      &    42.70     &     53.20  &  47.95  \\
%		CSI-GDAM~\cite{zhang2021csi}           &   50.90     &   57.10  & 54.00   \\
%		AutoFi                  & \textbf{55.60} & \textbf{63.80} & \textbf{59.70}\\ \bottomrule
%	\end{tabular}
%\end{table}

\subsection{Ablation Study}
\begin{figure}[t]
	\centering
	\includegraphics[width=0.9\linewidth]{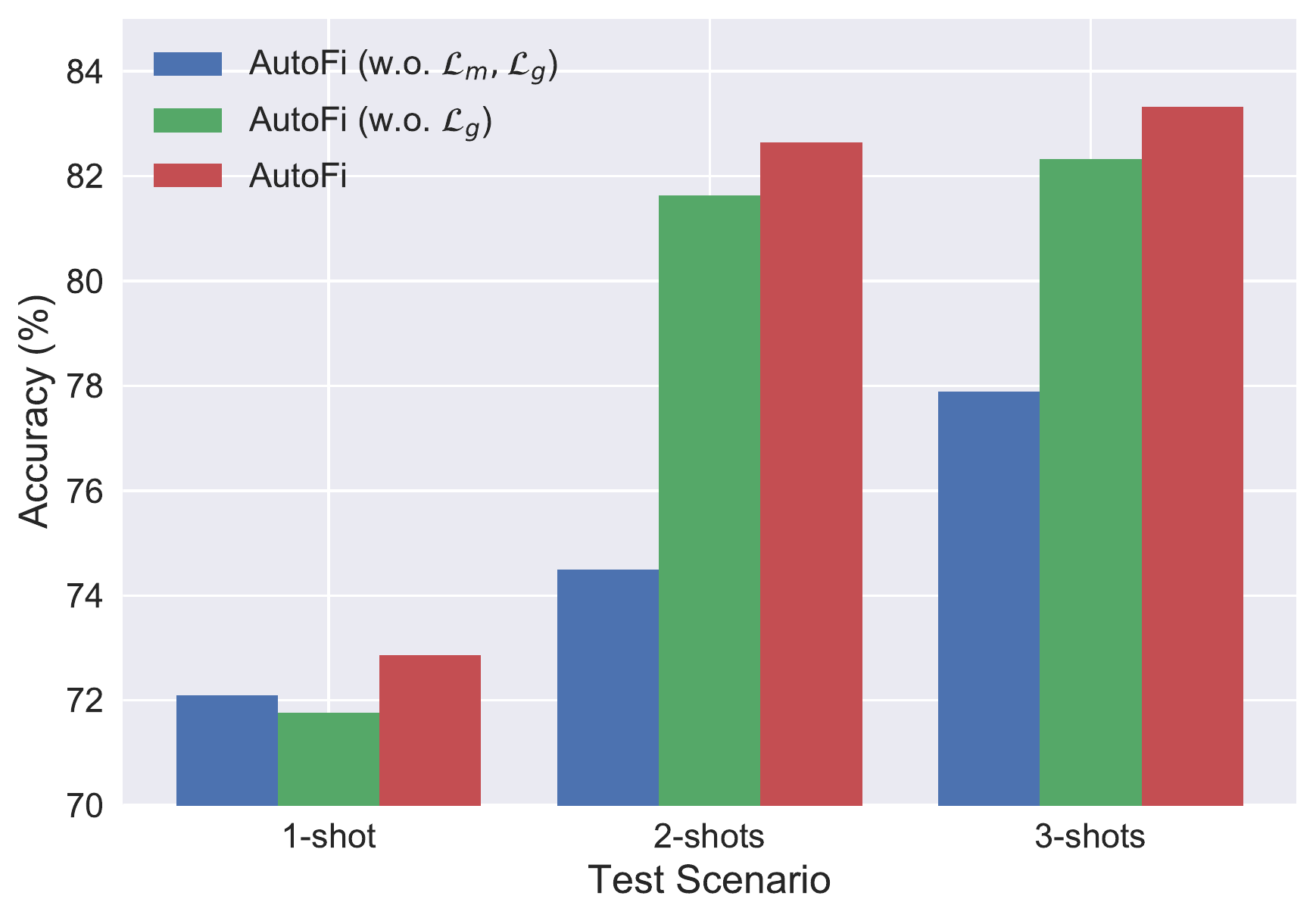}
	\caption{The ablation study of the proposed AutoFi method. The ``AutoFi (w.o. $\mathcal{L}_m,\mathcal{L}_g$)'' means that the GSS module of the AutoFi does not apply $\mathcal{L}_m$ and $\mathcal{L}_g$.}
	\label{fig:ablation}
\end{figure}

To demonstrate the effectiveness of multiple objectives in the GSS module, we compare our method with the cases of the lack of the mutual information loss and the geometric consistency. The baseline performance has been illustrated in the Table~\ref{tab:exp-real-world}, i.e. the prototypical network. Based on the real-world human identification experiments, we draw the results in Figure~\ref{fig:ablation}. The ``w.o.'' denotes ``without''. When the mutual information loss is absent, we can observe obvious performance decreasing for 2-shots and 3-shots cases. For 1-shot case, the performances are quite similar, because the scale of the training samples is rather limited. As for the geometric consistency, it leads to a marginal improvement for all scenarios, verifying its advantages for few-shot learning.

\subsection{Model Inference and Time Cost}
Though the AutoFi can learn CSI patterns by itself, model learning still requires computational resources. Here we compute the training time and the model inference time for each CSI sample. In a large-scale WiFi sensing, these data can be easily uploaded and processed at the cloud, so we run all the program on a single NVIDIA RTX 2080Ti. For our real-time system, the GSS module learns 5000 CSI samples for 300 epochs, which cost 22 mins. The FSC module only takes less than 1 minute. As this process is conducted offline, it is acceptable in reality. Compared to the model training, we pay more attention to the model inference for real-time systems. Our recognition model only costs 22ms for one CSI sample in our system. For UT-HAR and Widar, as the data dimensions are lower, the cost time is only 16ms and 15ms, respectively. In this manner, we prove that the AutoFi can be easily setup and run efficiently in the real world.

\section{Conclusion}\label{sec:conclusion}
In this paper, we propose AutoFi, a novel geometric self-supervised learning framework, which is the first work that realizes self-driven initialization of learning-based models using randomly-collected CSI data. The geometric self-supervised learning enables the AutoFi to learn CSI patterns by consistency and mutual information, and a few-shot calibration module can efficiently empower the AutoFi to conduct downstream recognition tasks. Extensive experiments are conducted in both real world and public datasets. The experimental results show that the AutoFi can significantly improve the few-shot performance, or enhance the existing systems by cross-task knowledge transfer. We believe that the AutoFi is an important step toward automatic and pervasive WiFi sensing. Future works may focus on how to leverage limited labeled samples by exploiting data augmentation and how to integrate WiFi and other modalities for robust sensing~\cite{zou2019wifi,deng2022gaitfi}.

\bibliographystyle{IEEEtran}
\bibliography{egbib}
\begin{IEEEbiography}[{\includegraphics[width=1in,clip,keepaspectratio]{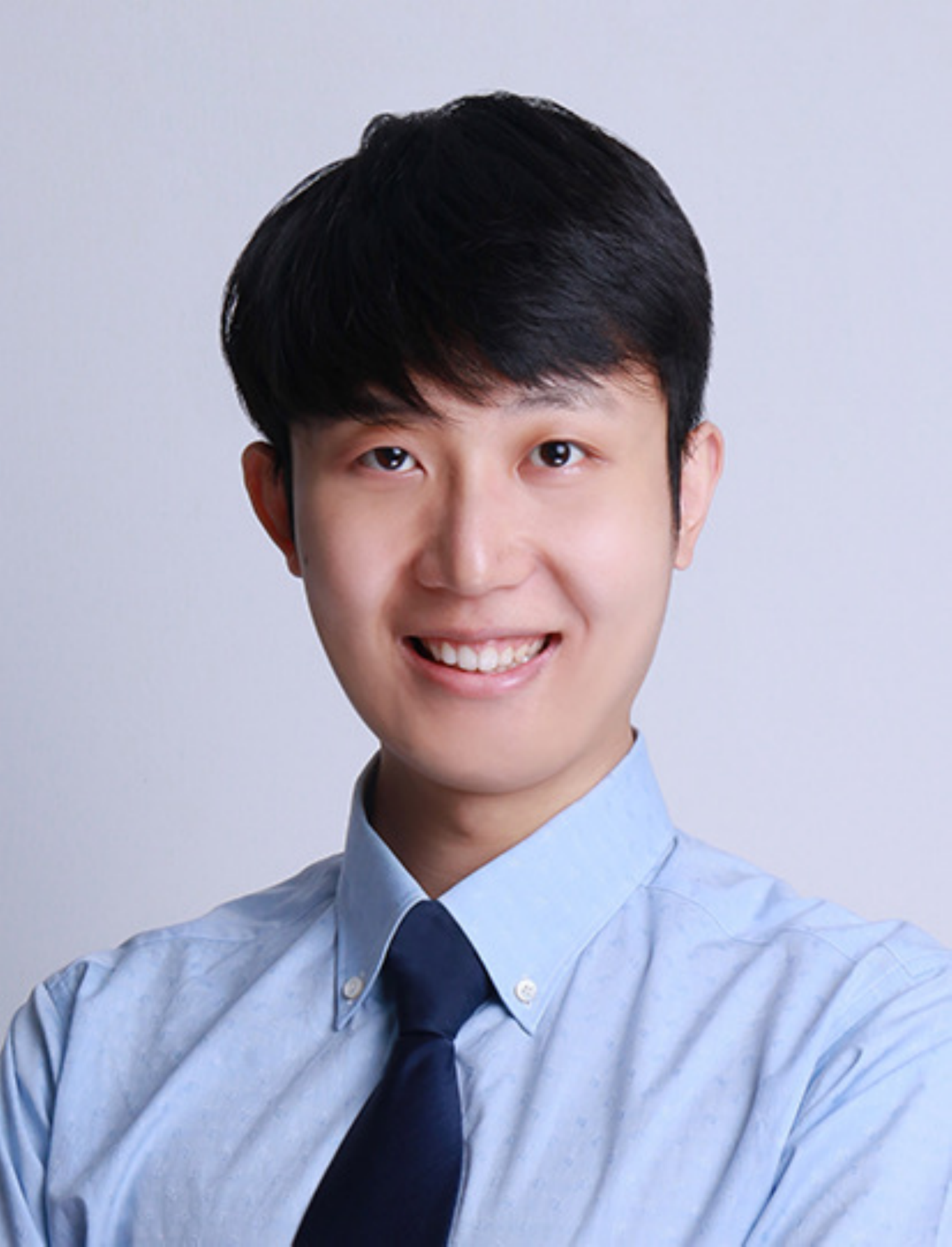}}]{Jianfei Yang} received the B.Eng. and Ph.D. from the School of Data and Computer Science, Sun Yat-sen University in 2016, and Nanyang Technological University (NTU), Singapore in 2021. He used to work as a senior research engineer at BEARS, the University of California, Berkeley. His research interests include deep transfer learning with applications in Internet of Things and computer vision. He won many AI and data challenges in the visual and interdisciplinary fields, such as ACM ICMI EmotiW-18, IEEE CVPR-19 UG2+ challenge and ICCV-21 Masked Face Recognition challenge. Currently, he is an independent principal investigator and a Presidential Postdoctoral Research Fellow at NTU.
\end{IEEEbiography}

\begin{IEEEbiography}[{\includegraphics[width=1in,height=1.25in,clip,keepaspectratio]{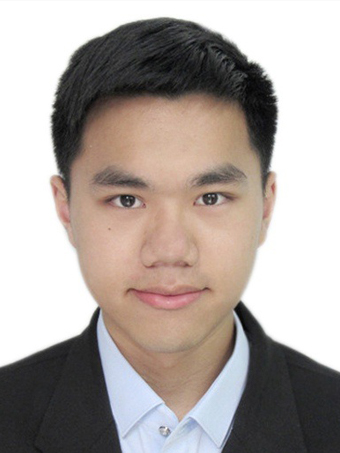}}]{Xinyan Chen} is currently an undergraduate student from the School of Electrical and Electronic Engineering, Nanyang Technological University, Singapore, and he worked for his Undergraduate Research Experience on Campus (URECA) program under the supervision of Prof Lihua Xie and Dr. Jianfei Yang at NTU. His research interests include deep learning and computer vision.
\end{IEEEbiography}

\begin{IEEEbiography}[{\includegraphics[width=1in,height=1.25in,clip,keepaspectratio]{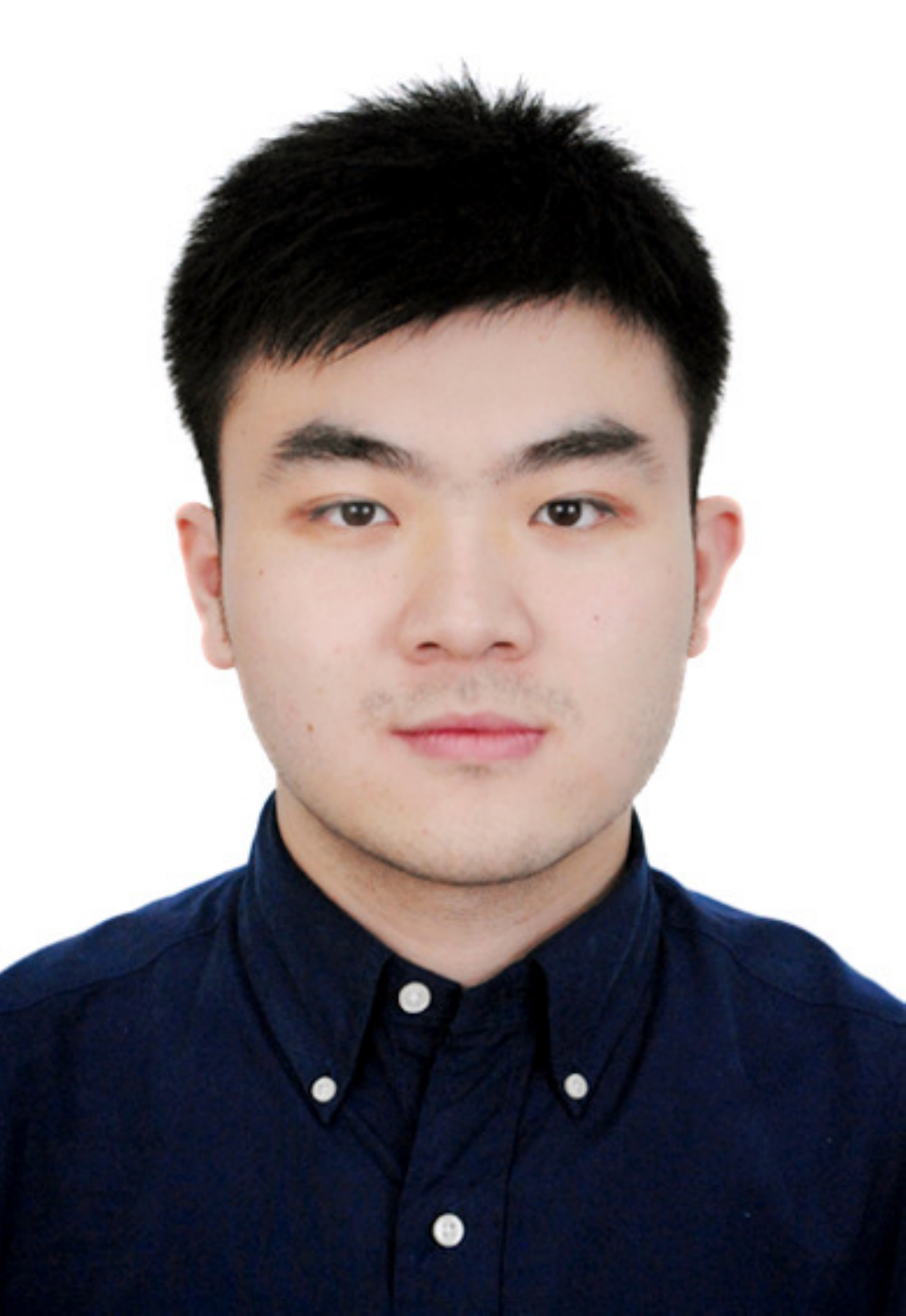}}]{Han Zou}
	received the B.Eng. (First Class Honors) and Ph.D. degrees in Electrical and Electronic Engineering from the Nanyang Technological University, Singapore, in 2012 and 2016, respectively. He is currently a Postdoctoral Scholar with the Department of Electrical Engineering and Computer Sciences at the University of California, Berkeley, CA, USA. His research interests include ubiquitous computing, statistical learning, signal processing and data analytics with applications in occupancy sensing, indoor localization, smart buildings and Internet of Things.	
\end{IEEEbiography}

\begin{IEEEbiography}[{\includegraphics[width=1in,height=1.25in,clip,keepaspectratio]{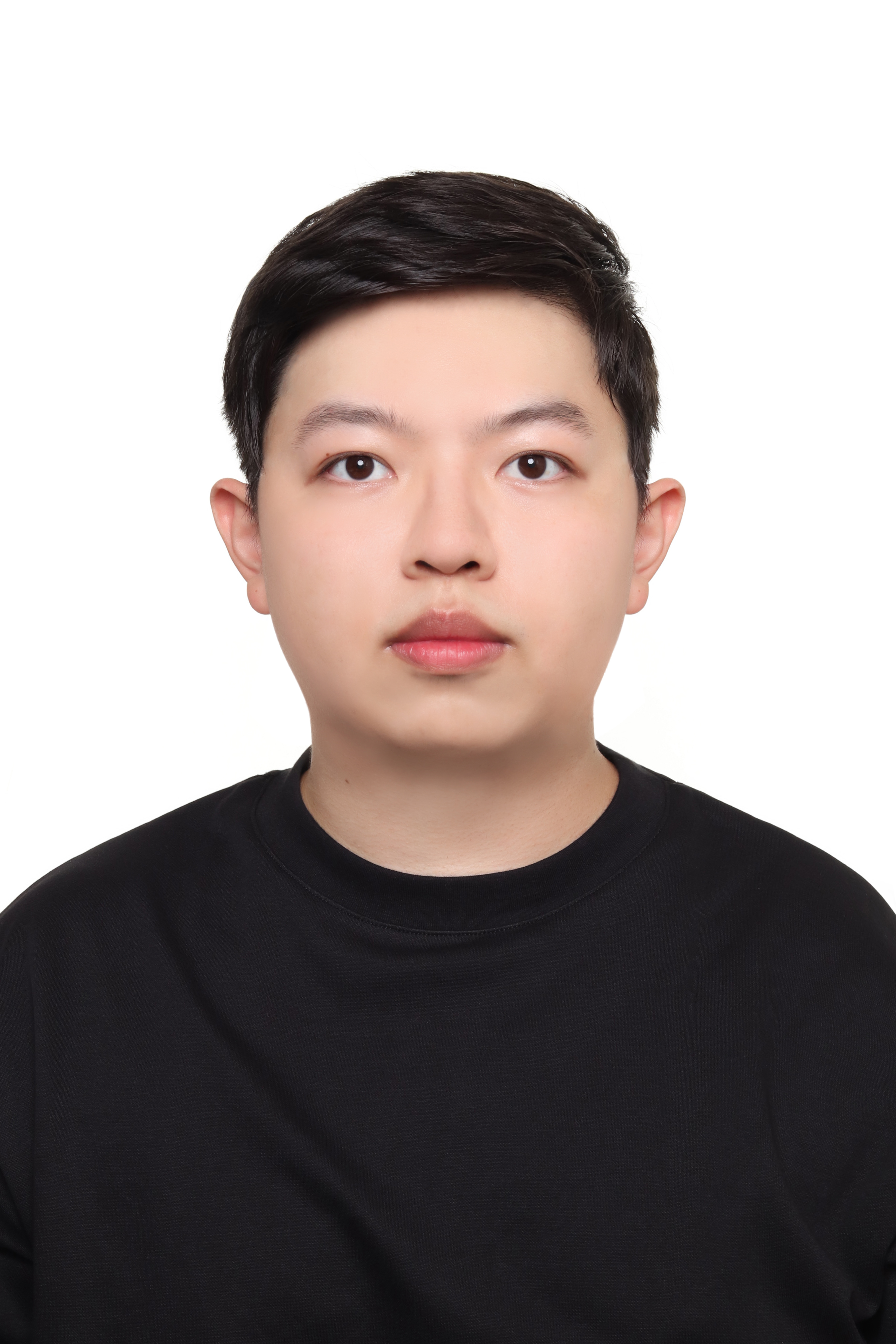}}]{Dazhuo Wang}
	received the B.Eng. from the School of Electrical and Electronic Engineering, Nanyang Technological University in 2018. He is currently a PhD candidate in the School of Electrical and Electronic Engineering, Nanyang Technological University, Singapore. His research interests include Industrial Internet of Things and machine learning. He is a scholar of Agency for Science, Technology and Research (Singapore) under AGS scholarship.
\end{IEEEbiography}

\begin{IEEEbiography}[{\includegraphics[width=1in,height=1.25in,clip,keepaspectratio]{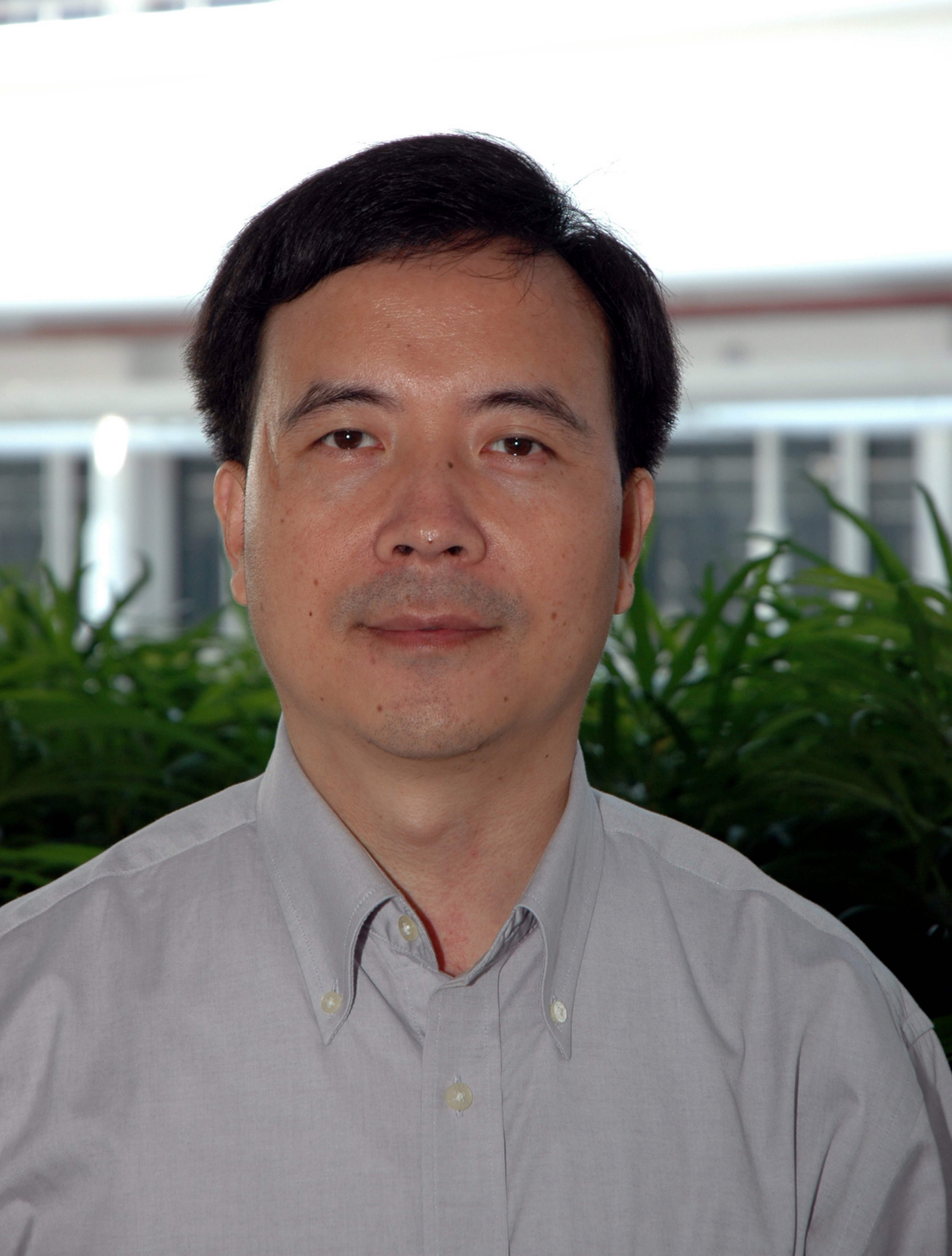}}]{Lihua Xie}
	received the Ph.D. degree in electrical engineering from the University of Newcastle, Australia, in 1992. Since 1992, he has been with the School of Electrical and Electronic Engineering, Nanyang Technological University, Singapore, where he is currently a professor and Director, Center for Advanced Robotics Technology Innovation. He served as the Head of Division of Control and Instrumentation and Co-Director, Delta-NTU Corporate Lab for Cyber-Physical Systems. He held teaching appointments in the Department of Automatic Control, Nanjing University of Science and Technology from 1986 to 1989. 
	
	Dr Xie’s research interests include robust control and estimation, networked control systems, multi-agent networks, smart sensing and unmanned systems. He is an Editor-in-Chief for Unmanned Systems and has served as Editor of IET Book Series in Control and Associate Editor of a number of journals including IEEE Transactions on Automatic Control, Automatica, IEEE Transactions on Control Systems Technology, IEEE Transactions on Network Control Systems, and IEEE Transactions on Circuits and Systems-II. He was an IEEE Distinguished Lecturer (Jan 2012 – Dec 2014). Dr Xie is Fellow of Academy of Engineering Singapore, IEEE, IFAC, and CAA.
\end{IEEEbiography}

\end{document}